\documentclass{raa}

\usepackage{graphicx, times}                
\usepackage{natbib}
\usepackage{amssymb,amsmath}
\bibpunct{(}{)}{;}{a}{}{,}

\usepackage[a4paper=true,dvipdfm=true,pagebackref=true]{hyperref}
\hypersetup{colorlinks = true, linkcolor = green, anchorcolor = red, citecolor = blue, filecolor = red, pagecolor = red, urlcolor = red}

\begin{document}

   \title{Study of variable stars associated with maser sources:  
}
  \subtitle{G025.65+1.05}

   \volnopage{Vol.0 (20xx) No.0, 000--000}      
   \setcounter{page}{1}          

   \author{A. M. Sobolev
      \inst{1}
   \and A. P. Bisyarina
      \inst{1}
   \and S. Yu. Gorda
      \inst{1}
   \and A. M. Tatarnikov
      \inst{2}
   }

   \institute{Astronomical Observatory of the Ural Federal University 
             Ekaterinburg 620000, Russia; {\it Andrej.Sobolev@urfu.ru}\\
        \and
             Sternberg Astronomical Institute of Moscow State University\\}
      
\vs\no
   {\small Received~~2018 April 30; accepted~~2018~~November 07}

\abstract{We report variation of K-band infrared (IR) emission in the vicinity of the G025.65+1.05 water and methanol maser source. New observational data were obtained with 2.5m~telescope of the Caucasian Mountain Observatory (CMO) of Moscow State University on~2017-09-21 during the strong water maser flare. We found that the IR source situated close to the maser position had decreased brightness in comparison to archive data. This source is associated with a massive young stellar object (MYSO) corresponding to the compact infrared source IRAS~18316-0602 (RAFGL~7009S). Similar decrease in K-brightness of the IR source close to the maser position was observed in March~2011 when the water maser activity was increased. The dips in MYSO brightness can be related to the maser flare phases. Maser flares that are concurrent with dips of the IR emission can be explained if the lower IR radiation field enables more efficient sink of the pumping cycle by allowing IR photons to escape the maser region. 
}

   \titlerunning{Study of stars with maser sources: G025.65+1.05}  

   \maketitle

\section{Introduction}           
\label{sect:intro}

Significant increase of the infrared (IR) flux at the vicinity of flaring methanol masers were recently reported for NGC~6334I~MM1 (\citealt{Hunter+etal+2017}), S255 (\citealt{Stecklum+etal+2016, Zinchenko+etal+2017}), and G107.298+5.639 (\citealt{Stecklum+etal+2018}). 
Noteworthy, in NGC~6334I high activity of the water and methanol masers was contemporaneous (\citealt{Hunter+etal+2017, MacLeod+etal+2018}). But the Very Large Array observations of the water masers (\citealt{Brogan+etal+2018}) showed that the majority of the flaring water maser emission originated from the synchrotron source north along the jet driven by the source MM1 while the water maser emission toward the flaring IR source MM1 dropped. Meanwhile, the water and methanol maser flares in the G107.298+5.639 were alternating (\citealt{Szymczak+etal+2016}). Further, no significant flares or dimming of emission were reported for the water maser in S255 during the strong methanol maser flare which took place in~2015 and~2016 (\citealt{Fujisawa+etal+2015, Szymczak+etal+2018}). This shows that association between flares in IR continuum and in maser lines of different molecules can have different nature. 

In this paper, we consider near-IR variability in the vicinity of the water maser source G025.65+1.05 which recently experienced strong flares (\citealt{Lekht+etal+2018, Volvach+Volvach+etal+2017a, Volvach+Volvach+etal+2017b, Ashimbaeva+etal+2017}). The vicinity of this maser contains the compact infrared source IRAS~18316-0602 (RAFGL 7009S) with luminosity of about~{$3{\times}10^4 L_{\odot}$} (\citealt{McCutcheon+etal+1995}) and the ultracompact H{\small{II}}~region G025.65+1.05. The radio source, first identified at~3.6~cm by \cite{Kurtz+etal+1994}, coincides spatially with submillimeter emission at $350~\mu$m (\citealt{Hunter+etal+2000}), $450$ and~$850~\mu$m (\citealt{Walsh+etal+2003}). The region contains a massive young stellar object (YSO) which drives CO bipolar outflow (\citealt{Shepherd+Churchwell+1996}). \cite{Zavagno+etal+2002} suggested that RAFGL~7009S is an embedded young stellar object ``associated with the ultracompact H{\small{II}} region G025.65+1.05, which may be excited by a B1V~star". Kinematic distance estimation on the basis of ammonia line observations gives the value of about~3.2~kpc (e.g. \citealt{Molinari+etal+1996}).

A prominent bright rapid flare of the main feature of the water maser took place in September~2017  -- flux density raised from less than 1~kJy to about 20~kJy in a few days (\citealt{Volvach+Volvach+etal+2017a}). This flare and the previous one were preceded by a moderate rise of the methanol maser emission which happened 3~months in advance of the water maser flare (\citealt{Sugiyama+etal+2017}). There were two more bright flares in October-November  (\citealt{Volvach+Volvach+etal+2017b, Ashimbaeva+etal+2017}). The latter one lasted only for a couple of days, reached 76~kJy  at the maximum and faded to 16~kJy within a day (\citealt{Ashimbaeva+etal+2017}). 

We obtained K-band data of G025.65+1.05 on~2017-09-21 (soon after the peak of the first maser flare) at the Caucasian Mountain Observatory (CMO),  Sternberg Astronomical Institute of Moscow State University. Variation of G025.65+1.05 K-band flux density distribution was noticed ``by eye" after comparison with an archive UKIDSS image (\citealt{Sobolev+etal+2017}). First photometric data were given in \cite{Stecklum+etal+2017}. In this paper, we present a new photometric study of G025.65+1.05 IR-variability.

\section{CMO observations}
\label{sect:Obs}

Observations of G025.65+1.05 were obtained at CMO on~2017-09-21 in the infrared K-band using ASTRONIRCAM (\citealt{Nadjip+etal+2017}). The instrumental photometric system is close to the standard MKO (Mauna Kea Observatories) photometric system. Camera was set in the imaging mode. The final image is the sum of 50 separate images obtained with an exposure of 3.67~seconds with 3~arcsec dithering. For these separate images, the bias, dark and flatfield correction were conducted. The FWHM (CMO point-spread-function) of the resulting image is  about~1.1~arcsec.

At the date of the CMO observations, the flux density of the water maser was about 15~kJy (Volvach~A.~E., private communication). We also used archival IR data and data of previous observations from the literature. Data and  references are listed in Table~1.

\begin{table}
\bc
\begin{minipage}[]{100mm}
\caption[]{Observational Data\label{tab1}}\end{minipage}
\setlength{\tabcolsep}{5pt}
\small
 \begin{tabular}{cccccc}
  \hline\noalign{\smallskip}
Date & MJD & Instrument & Filter &$\lambda_{central},$ & $\Delta\lambda$,\\
(YYYY-MM-DD)& & & & ${\mu}m$ &  ${\mu}m$\\
  \hline\noalign{\smallskip}
2003-06-16& 52806.4 &UFTI, UKIRT &K&2.20&0.34\\
2004-07-11 & 53197.7 &IRIS2, AAT &K$_s$&2.14&0.32\\
2007-08-27 & 54339.3 &WFCAM, UKIRT &K&2.20&0.34\\
2011-03-20 & 55640.4 &OSIRIS, SOAR &C2&2.14&0.05\\
2011-09-18 & 55822.3 &WFCAM, UKIRT&K&2.20&0.34\\
2017-09-21 & 58017.6 &CMO & K&2.19&0.32\\
  \noalign{\smallskip}\hline
\end{tabular}
\ec
\tablecomments{0.86\textwidth}{UFTI data is reported in \cite{Varricatt+etal+2010}, AAT data in \cite{Longmore+etal+2006} and SOAR data in \cite{Navarete+etal+2015}}
\end{table}

\section{Data reduction}
\label{sect:data}

\begin{figure}
   \centering
  \includegraphics[width=14.5cm, angle=0]{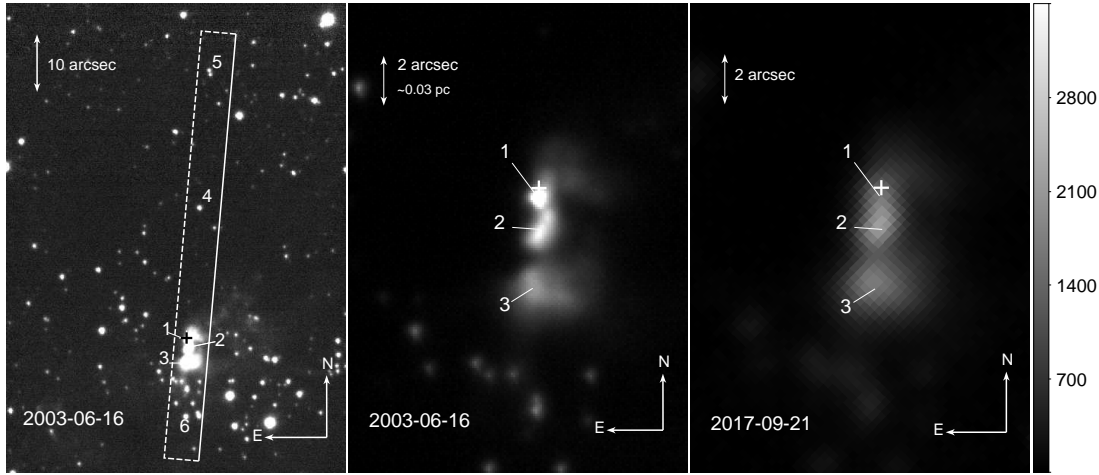}
   \caption{Vicinity of G025.65+1.05 in K-band. The water maser position from \cite{Jenness+etal+1995} is marked by a cross. The rectangle in the left panel shows the region in which flux density distribution is measured. The numbers~1 and~2 indicate the peak position of the IR sources nearest to the water maser position. Star~4 was used for calibration, stars 5 and 6 for calibration control. The two left panels show data obtained on 2003-06-16 in different spatial and brightness scales. The spatial scale is given at the top of each panel, the distance from the source is taken from \cite{Molinari+etal+1996}. The right panel shows the CMO data with the same brightness scale as the central image.  The numerical scale in units $10^{-18} W/m^2/\text{micron}/\text{arcsec}^2$ is based on 2MASS catalog value for star~4. The marked positions of~1-3 sources in the right panel are  taken from  peak positions of the central image. The flux density decrease of the source~1 in~2017 is clearly seen.}
   \label{Fig1}
   \end{figure}

We analyzed the flux density distribution along a chosen line passing through the considered source. The flux density was integrated in the tangential direction within the breadth of considered rectangle with the size of 75~arcsec~$\times$~6~arcsec (shown in Figure~\ref{Fig1}). It contains three IR sources close to the maser position and three isolated stars used for calibration and calibration control. The rectangle breadth was chosen in order to cover at least 90~per cent of emission of these objects. We calibrated the data to uniform scale assuming that the star marked~4 does not vary. Photometric analysis had shown that this star varied only within the limits of 0.1~mag from its mean value for all our data. Extreme variations of this star were the following: magnitude of star~4 increased by about 0.1~mag on~2007-08-27 and decreased on~2011-09-17 by about the same value. Images in these dates were obtained by UKIDSS. So, we used the average signal to get the UKIDSS data calibration parameters.

The numerical scale of obtained flux density was derived from the 2MASS K-flux for star~4. The calibration coefficients  were found by fitting of the integrated  observed signal of  star~4 and it absolute flux density,  calculated from the 2MASS point source catalog value $K=14.525\text{ mag}$. To facilitate comparison of the data we smoothed images in order to have the same angular resolution as in CMO observations (convolution process). 
The convolution was conducted  with the IRAF software. The resulting flux density profiles are shown in Figures~\ref{Fig2} and~\ref{Fig3}. We did not attempt to obtain exact absolute values of the flux densities because further considerations are based on analysis of the relative flux density changes. 

\section{Results and discussion}
\label{sect:discussion}
\begin{figure}
   \centering
  \includegraphics[width=13cm, angle=0]{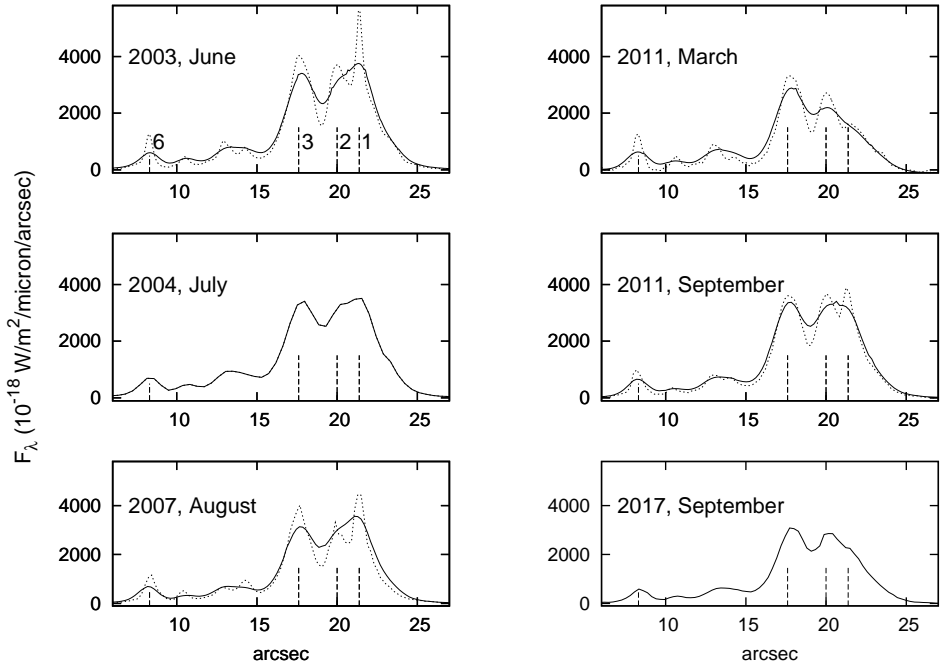}
   \caption{The flux density distribution along the rectangle shown in Figure~\ref{Fig1}. The vertical scale is rough and was obtained from $K=14.525\text{ mag}$ 2MASS catalog value for star~4. The horizontal scale shows the angular distance (in arcseconds) from (J2000):  $\text{RA}=18^\text{h}34^\text{m}20.952^\text{s}$ and $\text{DEC}=-5^\circ59'25''.375$. The solid line represents the data after smoothing to the CMO angular resolution, dotted line shows the data without angular smoothing (for the cases when the angular resolution of the data is different from the CMO data one). The dashed bar marked by~1 indicates peak position of the IR source nearest to the water maser. The positions of sources, marked by dashed bars in all panels, are taken from the peak positions of the first panel image (2003-06-16, data with better angular resolution). The source designations are the same as in the Figure~\ref{Fig1}. The K-band intensity of source~1 was significantly less in March~2011 and September~2017 with respect to other epochs.} 
   \label{Fig2}
   \end{figure}

\begin{figure}
   \centering
  \includegraphics[width=13cm, angle=0]{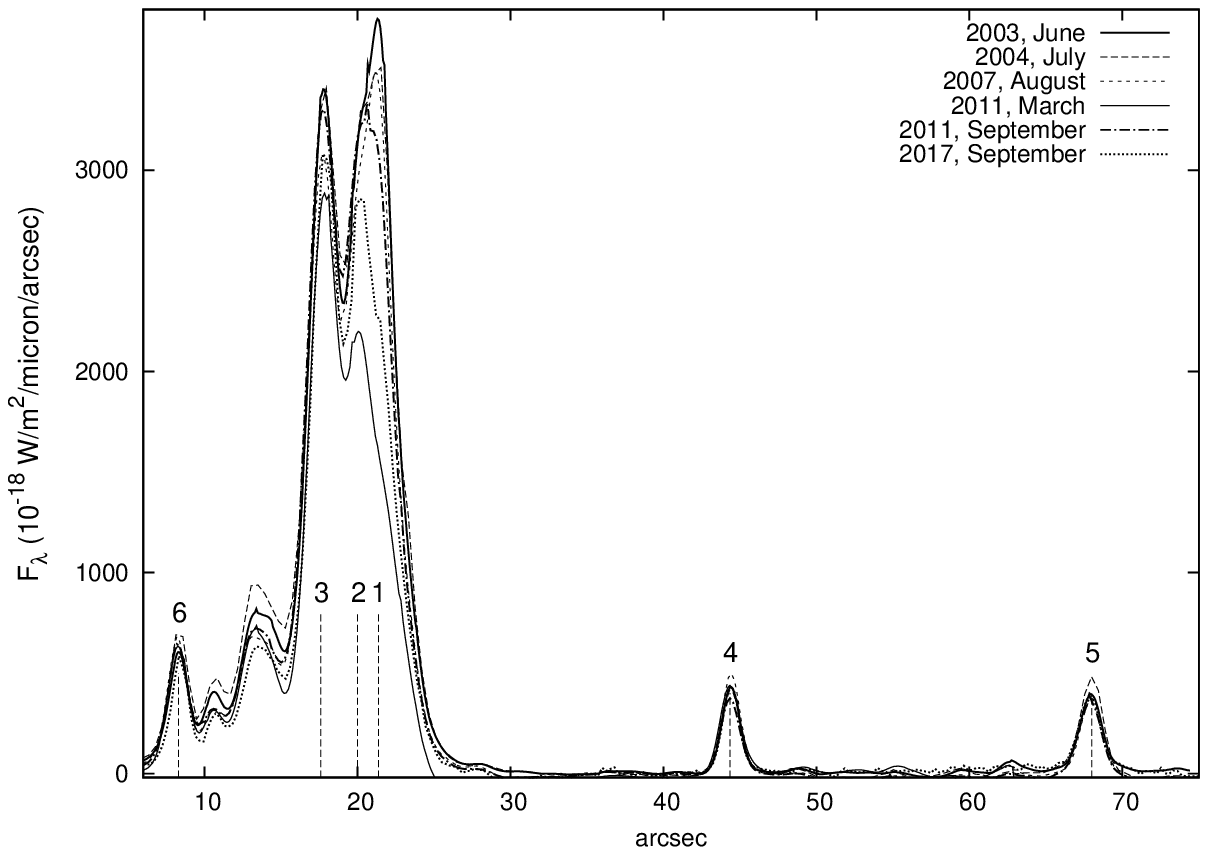}
   \caption{The superposition of the flux density profiles along the rectangle shown in Figure~\ref{Fig1}. The designations are similar to those in Figure~\ref{Fig1} and Figure~\ref{Fig2}. Star~4 was used for calibration and stars~5 and~6 for calibration control. All data smoothed to the CMO angular resolution. Source~1 was significantly fainter in March~2011 and September~2017 with respect to the other epochs.} 
   \label{Fig3}
   \end{figure}

Figures~\ref{Fig2} and \ref{Fig3} show that the IR source nearest to the maser position (marked by~1 in the figures) was significantly fainter in September~2017 and March~2011 in comparison to other epochs. Note that the image from March~2011 was obtained with OSIRIS/SOAR using a narrow continuum filter centered at~2.14 ${\mu}$m (C2~in Table~1) while the other images were obtained with broad K-band filters. In September~2011 the emission of the source was slightly dimmer  compared to the majority of the epochs. We think that the low flux density in the SOAR data is likely due to actual variability and not to calibration artifacts.

The epoch of our CMO observation is close to the maser flare. Consideration of the results of long-term water maser monitoring by \cite{Lekht+etal+2018} shows that the epochs of IR observations in~2011 correspond to the periods of  the increased maser activity (with several kJy in March and about 400~Jy in September).
The other epochs  presented in Table~1 correspond to a considerably lower state of maser activity  with flux densities less than 200~Jy. Therefore, our analysis shows a possible connection between water maser flares and dips in IR emission. 

A half a day rise and two days duration of the recent maser flare reported by \cite{Ashimbaeva+etal+2017} imply that the flare is caused by changes in the radiative part of the pumping process since the geometrical changes and collisional events are expected to be slower: even in the W49N with its extremely powerful (for the interstellar objects) shocks, the time duration of the fastest detected flare is considerably longer (\citealt{Liljestrom+Cwinn+2000}). Substantial influence of the radiative processes is very likely because they play a very important role in the water maser pumping (\citealt{Gray+2012, Gray+etal+2016}). Since water maser flares presumably happen during dips of the IR emission, it is not likely that the IR radiation makes the main contribution to the source of the maser pumping. This corresponds to results of theoretical considerations which show that the pumping of the strong water masers with the radiative source is unlikely (\citealt{Strelnitskii+1984, Deguchi+1981, Shmeld+1976}) because the pump power in these cases is insufficient. 
We have to note that an analysis of the role of IR radiation in the pumping without consideration of the maser sink is not complete. The sink of the pumping is often ignored but it can play a crucial role in the maser formation (\citealt{Strelnitskii+1981, Sobolev+Gray+2012}). For the water masers, theoretical considerations show that the sink of the IR photons due to the cold dust absorption within the masing region can give birth to the strong masers (for far IR photons sink \cite{Deguchi+1981}  and  for near IR photons sink \cite{Strelnitskii+1977}). Alternating character of the IR radiation density and water maser activity finds observational support in the drop of the water masers flux accompanied by the increase of the radiation field recently observed in the NGC~6334I-MM1 by \cite{Brogan+etal+2018}. 

An increase of the water maser sink efficiency is also realized in the case when the IR photons obtain possibility to escape from the masing region. This possibility is blocked when the radiation density in the masing region surroundings is high. Maser flares that are simultaneous with dips of the IR emission can be explained by the following effect: the decrease of the IR radiation field increases efficiency of the sink of the pumping mechanism by allowing more IR photons to escape from the masing region. 

At present, we don't have data describing changes of the near and far IR radiation density during the G025.65+1.05 maser flare which is necessary for the full analysis of the situation. Anyhow, low level of the K-band intensity of the source during the flare can indicate that the density of the radiation in the masing region is probably reduced at the longer IR wavelengths as well. Such a situation takes place in the intermediate mass YSO G107.298+5.639. Recent observations of this source by \cite{Stecklum+etal+2018} have shown that the periods of high water maser activity coincide with the periods when the near-IR K-band and mid-IR NEOWISE W1, W2, W3 and~W4 intensities of the source are reduced. 

\section{Conclusions}
\label{sect:conclusion}
We report a variability study of IR K-band flux density in the G025.65+1.05 vicinity. The IR source nearest to the water maser had significantly lower IR flux densities in March 2011 and September 2017 with respect to other considered epochs of observations. These two epochs are close to epochs when the water maser  was flaring. So, the K-band dips can have a relation to the water maser flares. We suggest that this relation is explained by the alternating character of the water maser sink efficiency and the IR radiation field density in the vicinity of the maser source.

\begin{acknowledgements}-
The authors are grateful to Navarete~F. for cooperation and providing the SOAR data and Volvach~A.~E. for the information about the maser activity. We thank the referee for helpful comments which allowed to increase quality of the paper. 
 A.~M.~Sobolev and S.~Yu.~Gorda were supported by the Russian Science Foundation grant 18-12-00193. A.~P.~Bisyarina was supported by Russian Foundation for Basic Research according to the research project 18-32-00314.
\end{acknowledgements}

\label{lastpage}

\end{document}